\documentclass[epj]{svjour}

\usepackage[english]{babel} 	
\usepackage{graphics} 
\usepackage{graphicx} 			
\usepackage{amssymb} 			
\usepackage{amsmath} 			
\usepackage{siunitx} 
\usepackage{subfig} 			
\usepackage{booktabs}

\usepackage{float}
\usepackage{url}

\begin{document}


\title{Assessing the accuracy of Hartree-Fock-Bogoliubov calculations by use of mass relations}
\author{D. Hove\thanks{e-mail: dennish@phys.au.dk}\and D.V. Fedorov\and A.S. Jensen\and K. Riisager\and N.T. Zinner}
\institute{Department of Physics and Astronomy, Aarhus University, DK-8000 Aarhus C, Denmark} 
\date{\today}


\abstract{
The accuracy of three different sets of Hartree-Fock-Bogoliubov calculations of nuclear binding energies is systematically evaluated. To emphasize minor fluctuations, a second order, four-point mass relation, which almost completely eliminates smooth aspects of the binding energy, is introduced. Applying this mass relation yields more scattered results for the calculated binding energies. By examining the Gaussian distributions of the non-smooth aspects which remain, structural differences can be detected between measured and calculated binding energies. Substructures in regions of rapidly changing deformation, specifically around $(N,Z)=(60,40)$ and $(90,60)$, are clearly seen for the measured values, but are missing from the calculations. A similar three-point mass relation is used to emphasize odd-even effects. A clear decrease with neutron excess is seen continuing outside the experimentally known region for the calculations. 
}

\maketitle


\section{Introduction}

The continuous increase in both quality and quantity of the experimental measurements of nuclear binding energies \cite{blo10} has also increased the expectations to the theoretical models. In spite of the many recent additions there are still thousands of unknown nuclei inside the driplines \cite{erl12}. Very accurate theoretical extrapolations are therefore needed, for instance, to predict the masses of the exotic elements involved in astrophysical processes. For such extrapolations to be reliable the model in question should reproduce the known nuclear masses on a global scale, both quantitatively and qualitatively.

Most current models are build on some variant of a Hartree-Fock-Bogoliubov (HFB) calculation \cite{lun03}. As a consequence we will be focusing on three different HFB calculations. Such self-consistent mean field theories are very well suited for global calculations, as their microscopic nature allows for the needed small scale individuality of the nuclei. Although the calculations are computationally rather more demanding than previous macro-microscopic models, it has been feasible with modern computers for more than a decade \cite{gor01}. 

One method for determining the consistency of nuclear binding energy calculations is to apply established mass relations to the calculations \cite{gar69,jen84,gar66}. Such mass relations rely both on the binding energy varying smoothly between neighbouring nuclei, and on other properties depending on the underlying principle behind the specific mass formula. For instance, the Garvey-Kelson (GK) mass relation \cite{gar66} is constructed to cancel the n-n, n-p, p-p interactions. Such mass relations are fulfilled surprisingly well for the measured binding energies, usually down to about $0.2$ MeV, whereas theoretical calculations, in particular HFB calculations, tend to fulfil such mass relations less consistently \cite{bar08}. As a result, mass relations can be useful in determining the validity of theoretical models. It has even been suggested that mass relations can be used to improve microscopic models \cite{bar08}. One of the main advantages of these mass relations is their flexibility, and many different forms exists. 

In this paper the focus will be on the use of simple three- and four-point mass relations \cite{jen84} to investigate the consistency of theoretical mean-field calculations of nuclear binding energies \cite{gor10,dob03,ber10}. These compact mass relations have the same main characteristic as the GK relations, that is cancellation of n-n, p-p, and n-p two-body interactions for smoothly varying, nuclear structures.

These mass relations were previously tested on the available experimentally obtained binding energies \cite{hov13-1,hov13-2}. In this paper the mass relations are applied to computed binding energies, where a much larger set of data, than the currently measured nuclei, is predicted. The compatibility or lack thereof should nevertheless be similar and thereby provide a test of the theoretical models.

In sect.~\ref{sec glo} the immediate differences between three different sets of calculated and the experimental measured binding energies are discussed, along with the differences between the three sets of calculations. In sect.~\ref{sec loc} a second order, four-point, mass relation is presented and applied. By cancelling almost all smooth and systematic effects, the minor fluctuations are highlighted. A three-point mass relation, designed to emphasize odd-even effects, is applied in sect.~\ref{sec oe}. Finally, in sect.~\ref{sec fluc} the results from applying the four-point mass relation are considered as Gaussian distributions. These distributions emphasize the structural differences between the measured and the calculated binding energies.

\section{Global differences \label{sec glo}}

Three sets of available, calculated binding energies are examined in detail. They are compared with the AME12 \cite{aud12} collection of measured binding energies.

The first set of calculations is based on the HFB approximation with generalized Skyrme forces and contact-pairing interaction. Included in the Skyrme force are two density-dependent generalizations of the $t_1$ and $t_2$ terms \cite{gor10}. The specific set examined here is the HFB27 mass excess calculations \cite{gorweb}, which includes both even and odd nuclei. These are converted to binding energies well-suited for our purposes. In the second set HFB calculations are performed, using less elaborate Skyrme forces, by expansion in a transformed harmonic oscillator basis. This is combined with the Lipkin-Nogami pairing method, and followed by particle-number projection \cite{dob03}. The SkP parameter set yields the lowest root mean square deviation compared with the experimental binding energies, and is the one examined here \cite{dobweb}. Unlike in the HFB27 calculations, only even-even nuclei are included in the SkP data set. However, calculated neutron and proton pairing gaps are also included, which is used in sect.~\ref{sec oe} to get an estimate of the odd-even effects. The third and final data set \cite{berweb} is from a constrained-HFB model using a D1S Gogny interaction (D1SG) \cite{ber10}. The mapping to the five dimensional collective Hamiltonian reflects the main purpose of studying excitations. Again only even-even nuclei are included in the data set, and a calculated pairing gap is not available.

\begin{figure}
\centering
\includegraphics[width=1.0\linewidth]{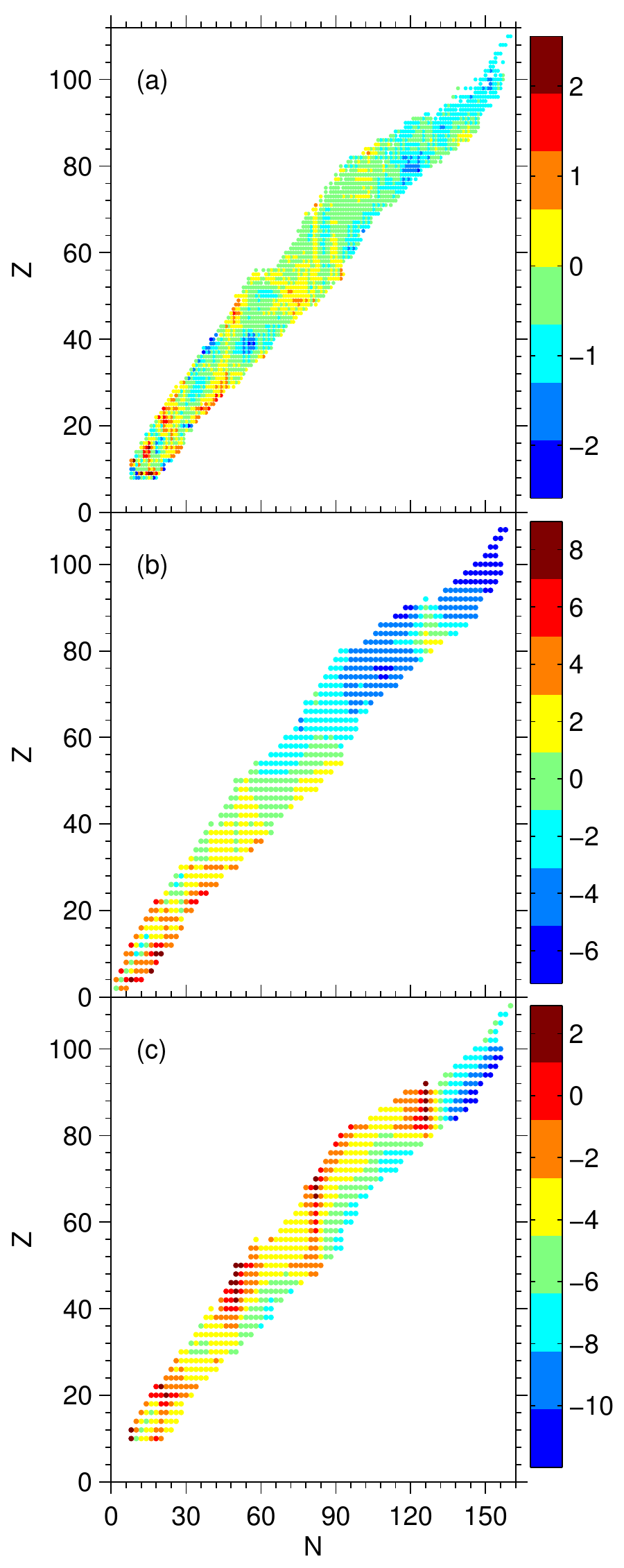}

\caption{(Colour online) Difference between the measured and calculated nuclear binding energy for (a) HFB27, (b) SkP and (c) D1SG. All energies are in MeV. \label{fig diff exp}}
\end{figure}

The differences between the calculated and the measured binding energies are shown in fig.~\ref{fig diff exp}. Figure \ref{fig diff exp}a shows the difference between the HFB27 results and the experimental binding energies for all known nuclei, $B_{\text{HFB27}} - B_{\text{Exp}}$. Overall, the deviations are very symmetric ranging from $-2.5 \, \si{\mega\electronvolt}$ to $2.5 \, \si{\mega\electronvolt}$. The greatest deviations are found among the light nuclei, which is not surprising as it is difficult to account for the very erratic nature of the very light nuclei in a global model. The shell patterns are not as prominent as one might expect, which indicates that the shell effects in the binding energies are well described by the model. The only exception is at $N=82$, where the difference is slightly positive in a region that is otherwise slightly negative. Approaching the super-heavy the difference becomes increasingly negative, although it is generally still only in the $1 \, \si{\mega\electronvolt}$ range. 

A greater deviation is found between the SkP results and the experimental values, presented in fig.~\ref{fig diff exp}b. This is only for even-even nuclei, as they are the only nuclei calculated with the SkP model. Here the difference, $B_{\text{SkP}}-B_{\text{Exp}}$, ranges from around $8 \, \si{\mega\electronvolt}$ for the very light to around $-6 \, \si{\mega\electronvolt}$ for the super heavy nuclei. The most visible shells are at $N=50$ and $N=126$, while the proton shells are almost undetectable, with the possible exception of $Z = 82$. Along the $N=Z$ line the absence of the Wigner effect results in SkP values being noticeably smaller than the experimental values.

Finally, in fig.~\ref{fig diff exp}c the difference between the D1SG results and the experimental binding energies, $B_{\text{D1SG}}-B_{\text{Exp}}$, is shown. Again, only even-even nuclei are calculated using this model. The difference tends to increase with distance to stability, from around $2 \, \si{\mega\electronvolt}$ at the proton rich side of stability, down to more than $-10 \, \si{\mega\electronvolt}$ for the super-heavy, neutron rich nuclei. More interesting is the fact that almost all the calculated values are smaller than the experimental measurements. The only exceptions lie along the magic numbers. As in fig.~\ref{fig diff exp}b the $N=Z$ line is also visible. This underestimate of the binding energies leaves room for neglected degrees of freedom.

\begin{figure}
\centering
\includegraphics[width=1.0\linewidth]{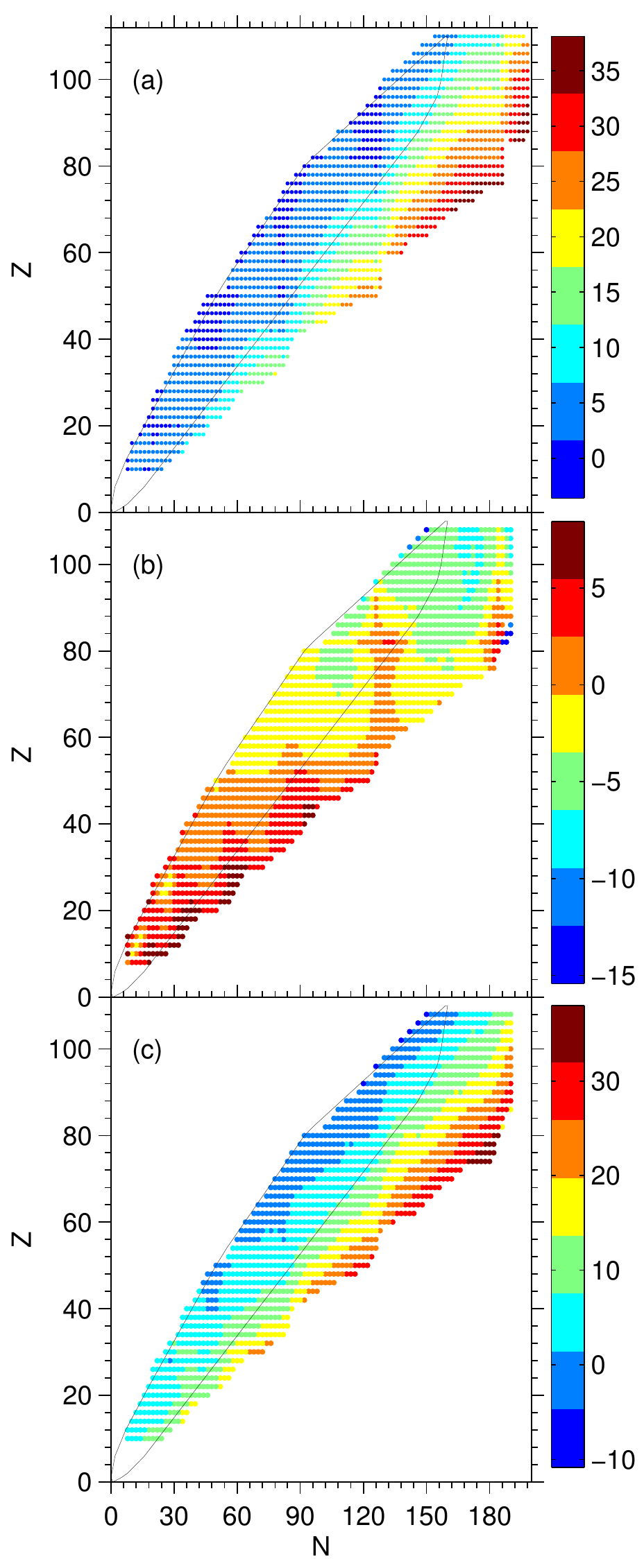}

\caption{(Colour online) The differences between the binding energies of the calculations (a) HFB27 and D1SG, (b) SkP and HFB27, and (c) SkP and D1SG. Included in black is a rough outline (the convex hull) of the known nuclei. All energies are in MeV. \label{fig diff ext}}
\end{figure}

Considering the difference is on the order of single MeV compared to the total value of several thousand MeV, the deviations are generally quite modest. Even though all three mean field models are based on some form of Hartree-Fock-Bogoliubov calculations, it is already apparent from fig.~\ref{fig diff exp} that there are significant differences between them. The differences between the three models are presented in fig.~\ref{fig diff ext}. 

The difference between the HFB27 and the D1SG models, $B_{\text{HFB27}} - B_{\text{D1SG}}$, is seen in fig.~\ref{fig diff ext}a. This demonstrates very clearly that even though the same data set form the basis for the models, and the models are similar in nature, very different results can be achieved. Only even-even nuclei could be compared, as the D1SG calculations only included those, but the calculations do extend far outside the experimentally known region. Outlined in black is the convex hull formed by the experimentally known nuclei. Unsurprisingly, the difference is rather modest inside this region, at most around $5 \, \si{\mega\electronvolt}$. However, this difference increases as the distance to stability increases, with the D1SG model consistently yielding smaller results than the HFB27 model. For the extremely heavy, neutron rich nuclei the difference is as great as $35 \, \si{\mega\electronvolt}$. Nuclei at the known magic numbers $20, 50, 82,$ and $126$ stand out in that the difference is less for these nuclei, indicating that the nuclear shells are treated similarly in the two models. Interestingly, another edge is seen at $N=188$, indicating a shell appears in the D1SG calculations, which is not present to the same extend in the HFB27 calculations.

A similar difference, only between the results of the SkP and the HFB27 model, $B_{\text{SkP}} - B_{\text{HFB27}}$, is seen in fig.~\ref{fig diff ext}b. An outline of the known nuclei is included as before. Again this is only for even-even nuclei, as the SkP model also only includes these nuclei. The difference changes from around $+7 \, \si{\mega\electronvolt}$ for the very light nuclei to around $-7 \, \si{\mega\electronvolt}$ for the super heavy, with a few nuclei deviating almost $-15 \, \si{\mega\electronvolt}$. Unlike previously, the difference does not change with distance to stability, instead it changes loosely with proton number. For very light nuclei the HFB27 calculations are consistently smaller than the SkP, at around $Z=50$ the two models agree very well, and for the super heavy nuclei the SkP values tend to be smallest. Once again the shells stand out, albeit differently than before. The SkP binding energies are slightly larger at, and directly following, neutron shells, while the proton shells are less pronounced. The HFB27 binding energies, on the other hand, tend to be slightly larger along $N=Z$, although the shells do counteract this tendency. A new neutron shell in the super heavy region also appears in fig.~\ref{fig diff ext}b, this time around $N = 184$.

Finally, the difference between the SkP and D1SG calculations, $B_{\text{SkP}}-B_{\text{D1SG}}$, is shown in fig.~\ref{fig diff ext}c. The same change with distance to stability as in fig.~\ref{fig diff ext}a is seen. This is not surprising as fig.~\ref{fig diff ext}c is just the sum of figs.~\ref{fig diff ext}a and \ref{fig diff ext}b. The difference then changes from around $-10 \, \si{\mega\electronvolt}$ for the heavy, proton rich nuclei to around $+35 \, \si{\mega\electronvolt}$ for the super heavy, neutron rich nuclei. As in fig.~\ref{fig diff ext}a the difference tends to be slightly less along magic numbers, indicating a greater agreement between the models along the nuclear shells.

The general scale of the deviations demonstrated by figs.~\ref{fig diff exp} and \ref{fig diff ext} is summarized in table \ref{tab rmse} by the root mean squared error values (rmse) 
\begin{align}
\text{rmse} = \sqrt{\frac{1}{n} \displaystyle\sum_{i=1}^n (x_i - \tilde{x}_i )^2}, \label{eq rmse}
\end{align}
where $n$ is the number of values, $x_i$ is the i'th value in the set, $\tilde{x}_i$ is the i'th reference value. This reference can either be one of the other extrapolations or the experimental values.

The rmse value is substantial between the D1SG and the HFB27 calculations, as well as between the D1SG and the SkP calculations, which is in agreement with fig.~\ref{fig diff ext}. On the other hand, there is a decent agreement between the HFB27 and the SkP calculations. The agreement is significantly better when comparing with the measured binding energies. The rmse values for the D1SG and the SkP calculations are $4.6$ and $3.1 \, \si{\mega\electronvolt}$ respectively, while the rmse value for the HFB27 calculation is as little as $0.7 \, \si{\mega\electronvolt}$. This is to be expected, as the HFB27 is the most elaborate model, involving the most parameters. The relatively large rms deviation of the SkP model's global calculations is recognized \cite{dob04}, but considering its simplicity the model is surprisingly accurate. The primary purpose of the D1SG model was not to calculate ground state energies, so a certain disparity is to be expected. It should also be noted that the HFB27 model is a more recent creation than both the D1SG and the SkP model. It is necessarily based on a larger set of measurements, and should provide a more accurate global fit. However, the HFB27 is substantially better and not far from the suggested chaotic limit of about $0.3$ MeV \cite{olo06}.

\begin{table}
\centering
\caption{The rmse values resulting from a comparison between the different data sets. The rmse values are calculated using eq.~(\ref{eq rmse}), and are in MeV. \label{tab rmse}}
\begin{tabular}{ l l *{1}{r}} 
\toprule
Model $(x_i)$               &  Reference $(\tilde{x}_i)$      & rmse     \\
\midrule
HFB27                       &       D1SG                      &     14.57                                       \\
HFB27                       &       SkP                       &     3.58                                        \\
SkP                         &       D1SG                      &     13.09                                       \\
HFB27                       &       Exp                       &     0.65                                        \\
D1SG                        &       Exp                       &     4.72                                        \\
SkP                         &       Exp                       &     3.17                                        \\
\bottomrule
\end{tabular}
\end{table}

\section{Local structures \label{sec loc}}

The specific variations in absolute values between the different models have been considered in the previous section. The present section will focus on the finer details of the structural differences. Of particular interest is the region outside the experimentally known nuclei. 

As the experimental measurements are extremely precise, a high level of accuracy should also be demanded of the theoretical models. Not only should a satisfactory model describe the global structures, it should also account for local variation and minor fluctuations. A method to eliminate the well known, large scale, smooth effects is needed to properly examine the minor fluctuations.

\subsection{The four point difference \label{sec 4 diff}}

The present technique eliminates smooth aspects of the binding energy up to and including second order by combining nuclei in a second order difference. The method is outlined below, for further details we refer the reader to \cite{jen84,hov13-1}. 

The binding energy is separated as
\begin{align}
B(N,Z) = \tilde{B}(N,Z) + \delta B(N,Z),
\end{align}
where $\tilde{B}$ contains all smooth parts, and $\delta B$ contains shell effects, and all other erratic or locally fluctuating parts. 

A second order difference given by
\begin{align}
Q(n_1,z_1;n_2,z_2) = &-S(N-n_1, Z-z_1) + 2S(N,Z) \notag \\ &- S(N+n_1, Z+z_1), \label{eq Q}
\end{align}
where $S$ is the separation energy
\begin{align}
S(N,Z) = B(N,Z) - B(N-n_2, Z-z_2),
\end{align}
would eliminate all smooth aspects up to and including second order in both $N$ and $Z$. The leading order contribution to the smooth part of $Q$ would then be
\begin{align}
\tilde{Q} 
= &-\frac{\partial^3 \tilde{B}}{\partial N^2 \partial Z}n_1(n_1 z_2+2n_2z_1) - \frac{\partial^3 \tilde{B}}{\partial N^3} n_1^2 n_2 \notag \\
&- \frac{\partial^3 \tilde{B}}{\partial N \partial Z^2} z_1(n_2z_1 + 2 n_1 z_2) - \frac{\partial^3 \tilde{B}}{\partial Z^3} z_1^2 z_2. \label{eq q rest}
\end{align}
The $n_i$ and $z_i$ values must be sufficiently small such that a structural similarity between the nuclei in question exist, otherwise a systematic cancellation cannot be expected. In the present, four-point case, $n_1=n_2=2$ and $z_1=z_2=0$ is chosen. The second order difference reduces to
\begin{align}
Q(2,0;2,0) &= B(N-4,Z)-3B(N-2,Z) \notag \\ &+3B(N,Z)-B(N+2,Z). \label{eq q 4}
\end{align}
This choice of a compact mass relation involve only four close-lying nuclei with the same proton number and either even or odd neutron number. A similar proton structure, which would be vertical in the (N,Z) plane of the chart of nuclei, can be created by choosing $n_1=n_2=0$ and $z_1=z_2=2$. The final structures used are
\begin{align}
\Delta_{2n} = \frac{1}{4} Q(2,0;2,0),&  &\Delta_{2p} = \frac{1}{4} Q(0,2;0,2), \label{eq del}
\end{align}
where the factor $1/4$ is to compensate for the combination of four nuclei. The $\Delta_{2n}$ structure is applied to the measured binding energies in fig.~\ref{fig D2n E}. In sect.~\ref{sec oe} a similar three-point mass relation is introduced to extract and examine odd-even effects.

These mass relations have the same main properties as the GK relations. For smoothly varying effects the third derivatives in eq.~(\ref{eq q rest}) are small, and the mass relation is very close to zero. As recently discussed \cite{hov13-1,hov13-2}, these mass relations are well suited for detecting and extracting structures rapidly changing with nucleon number.

The GK mass relations assume constant two-body interactions for all nuclei in the combination of binding energies. The number of pairs of identical interactions are then the same and cancel in these mass relations. Similar cancellation does not occur for three-body interactions in the GK mass relations. It is then tempting to consider whether signals of three-body interactions can be seen in our mass combination. We assume again that these interactions are constants, $C_3$, for all the four nuclei entering eq.~(\ref{eq q 4}).

It is then a matter of counting the number of similar three-body systems within these nuclei. We have four different combinations, that is three neutrons, three protons, two neutrons and one proton, or one neutron and two protons. The result is that all combinations cancel in the $\Delta_{2n}$ and $\Delta_{2p}$ mass relations, except for either the three neutrons or the three protons, which results in $-2 C^{(n)}_3$ and $-2C^{(p)}_3$ respectively. We emphasize that a similar calculation for two-body interactions gives zero.

Since the strengths decrease substantially from two to three-body interactions, we consider this to be a higher order contribution. It is contained in the third order derivative term in eq.~(\ref{eq q rest}) where contributions from all different types of nucleonic interactions are included. However, the nucleon number independent results of $-2$ times a smoothly varying constant could show up as a shift of the binding energy surfaces of the combinations in eq.~(\ref{eq del}).

In the more realistic case, where the strength parameters can change, the three-body contribution would not just be a constant displacement, but might very well fluctuate significantly. With varying strength parameters a larger contribution from lower order interactions would also be possible. In this case, the fluctuating nature of the contribution would likely be impossible to separate from the remains of the chaotic fluctuations of the binding energy.

\subsection{Application of the four point difference}

The most significant contributions, which remains when the four point structure is applied, come from the shell effects. A significant deviation should be visible when the structure actually crosses a major shell. The $\Delta_{2n}$ structure can then only detect neutron shells, when either $N$ or $N-2$ equals a magic number, while the proton shells are cancelled along with the smooth parts.

\begin{figure*}
\centering
\includegraphics[width=1.0\linewidth]{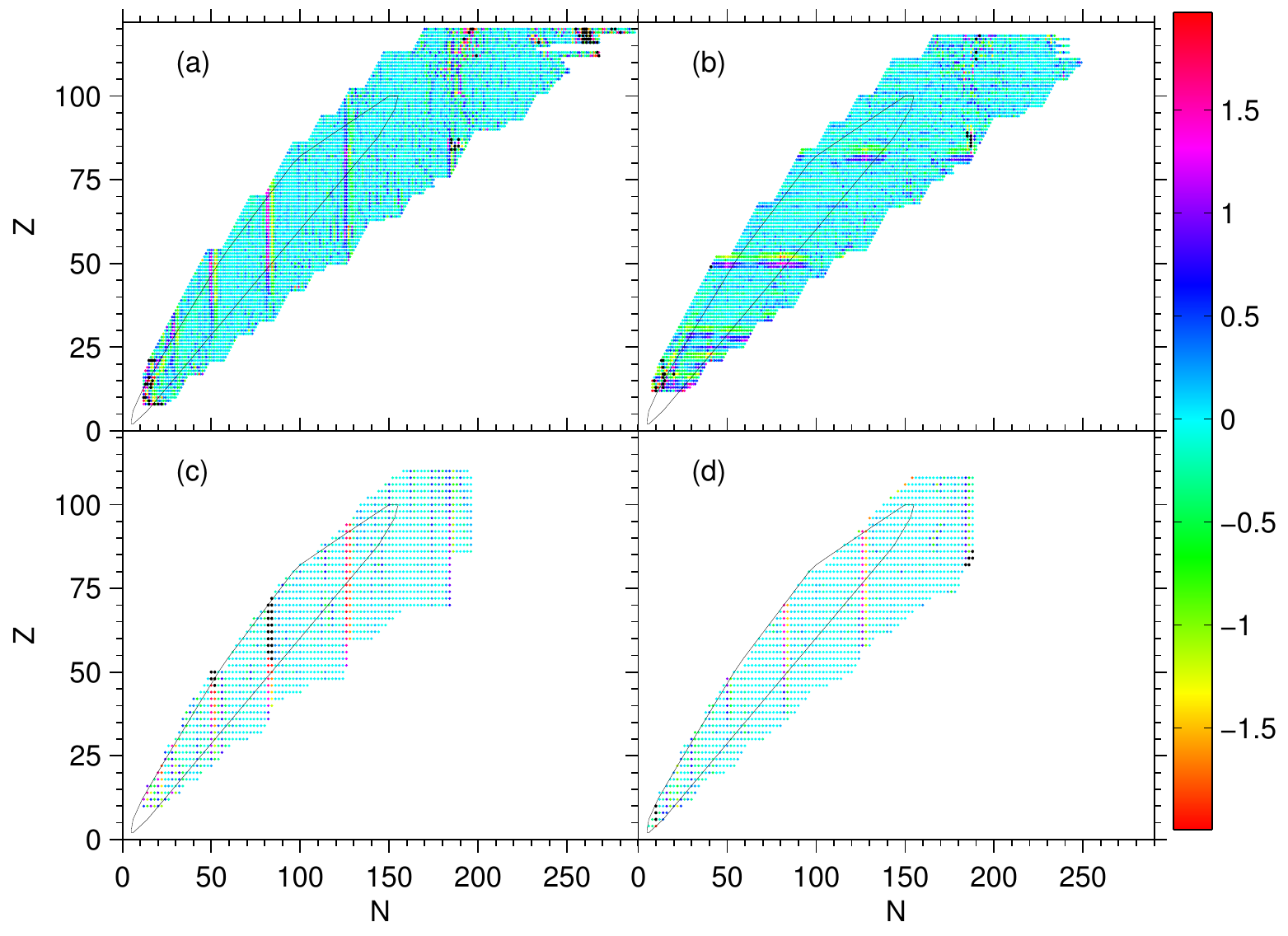}
\caption{(Colour online) (a) The $\Delta_{2n}$ mass relation applied to HFB27, (b) the $\Delta_{2p}$ mass relation applied to HFB27, (c) the $\Delta_{2n}$ mass relation applied to D1SG, (d) the $\Delta_{2n}$ mass relation applied to SkP. All energies are in MeV. Marked in black are the nuclei where $|\Delta|>2$ MeV. The experimental region is outlined. \label{fig D2n ext}}
\end{figure*}

The result of the four point neutron structure applied to the HFB27 calculations is seen in fig.~\ref{fig D2n ext}a. The same for the proton structure is seen in fig.~\ref{fig D2n ext}b. Nuclei for which $|\Delta| > 2$ MeV are marked in black. There are 57 such outliers for the $\Delta_{2n}$ structure and 24 for the $\Delta_{2p}$ structure, which deviates as much as $\pm 10 \, \si{\mega\electronvolt}$. Inside the experimental region the behaviour is as expected. The structure almost cancels everything, except for the shells, down to very small variations. The known nuclear shells continue outside the experimental region, although they tend to decrease in magnitude. Larger deviations are expected outside the experimental region, as mass formulas generally tend to obey Garvey-Kelson relations less outside the experimental region \cite{bar08}. A new, less pronounced, shell is seen at $N = 184$. Beyond the super heavy nuclei the fluctuations are seen to increase, and around $Z \sim 110$ and $N \sim 185$ and $N \sim 260$ something, which might be an island of stability, is seen. Most of the outliers are grouped in this region. Also just above the shell at $N=184$ for $Z=85-90$ a handful of nuclei deviate significantly from their neighbours. This is more clearly visible for the proton structure. In fig.~\ref{fig D2n ext}b the proton shells are slightly less pronounced than the neutron shells, but the general tendencies are the same.

In fig.~\ref{fig D2n ext}c the four point $\Delta_{2n}$ mass relation is applied to the D1SG calculations. Again the same shells are seen along with the shell at $N=184$, despite only even-even nuclei being included. There are 23 outliers for the D1SG calculation, deviating up to 3 MeV. The most significant difference between this and the other figures is the several minor, shell-like deviations seen in fig.~\ref{fig D2n ext}c. This is more pronounced for the super heavy nuclei outside the experimental region, but it is also seen around $N=114$ and $N=70$.

The result of the $\Delta_{2n}$ mass relation applied to the SkP calculations is seen in fig.~\ref{fig D2n ext}d. This includes 8 outliers, which deviates as much as $-11$ MeV. Even though only even-even nuclei have been calculated, many of the same tendencies are still seen. The neutron shells are very clearly seen inside the experimental region, but tend to decrease outside. Also a new shell is seen at $N = 184$. It is interesting to note that something similar to a very subtle shell structure begins at $(N,Z) = (166,86)$. However, this structure is not tied to a specific neutron number, but curves slightly. 

As in fig.~\ref{fig D2n ext}b the $\Delta_{2p}$ mass relation could be applied to the D1SG and the SkP calculations, but it does not reveal any new tendencies.

\begin{figure}
\centering
\includegraphics[width=1.0\linewidth]{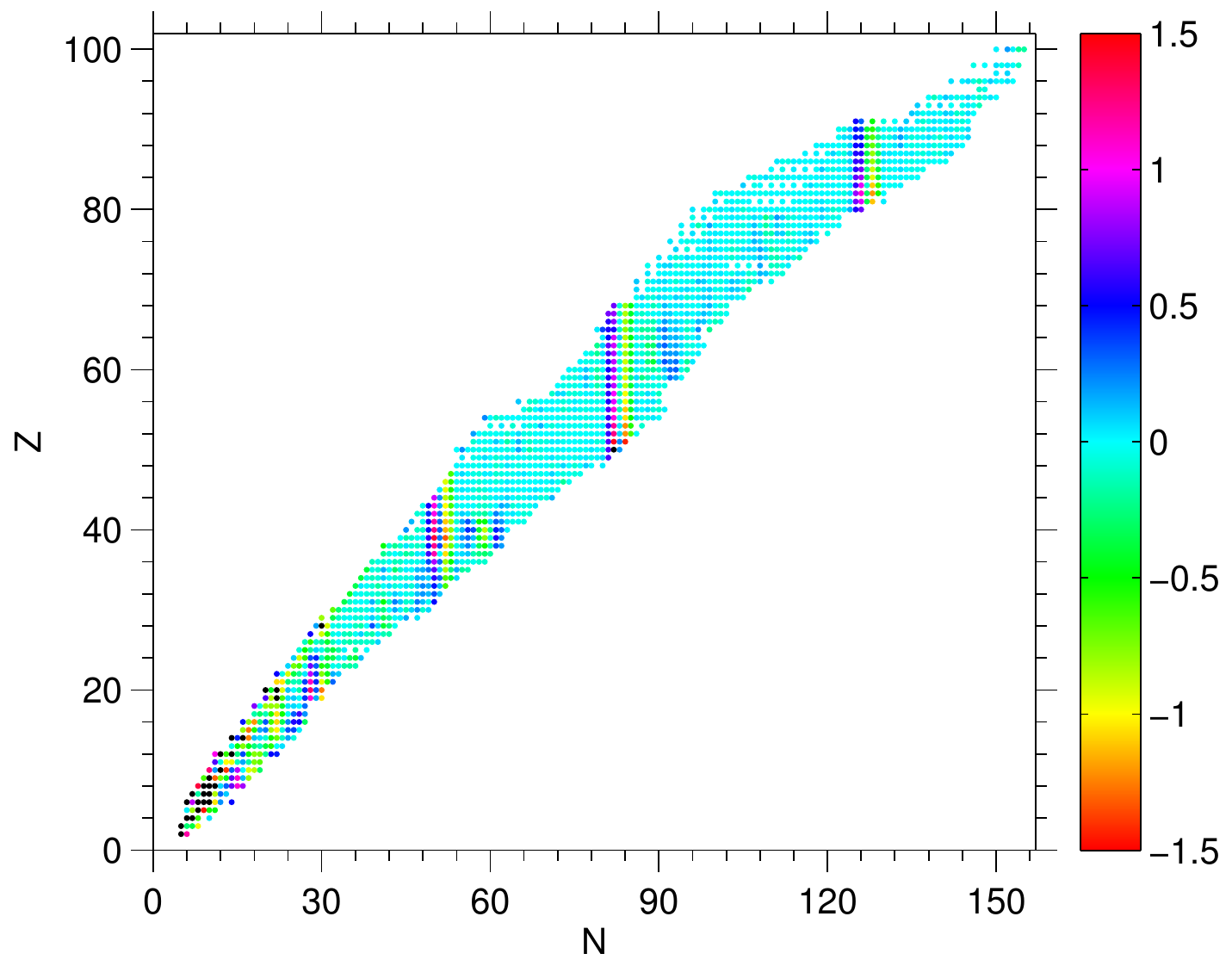}
\caption{(Colour online) The $\Delta_{2n}$ structure applied to the measured binding energies. The color scale is in MeV. Marked in black are the nuclei where $|\Delta_{2n}|>1.5$ MeV. \label{fig D2n E}}
\end{figure}

The four point neutron relation is also applied to the measured binding energies in fig.~\ref{fig D2n E}. Nuclei where $|\Delta_{2n}|>1.5$ MeV are marked in black. The 27 outliers are almost all among the very light nuclei, reflecting the fact that the variation between neighbouring nuclei is much greater for light nuclei, and the greatest deviation is $-5$ MeV. The known shells are clearly visible, but are significantly smaller in magnitude compared with fig.~\ref{fig D2n ext}. Focusing on nuclei with $A>50$ the largest deviation is around $1.5 \, \si{\mega\electronvolt}$ at the double magic $^{132}\text{Sn}$. However, even though the overall scale is smaller, there are still significant fluctuations between the shells. There is no sign of a constant displacement. All contributions from multi-nucleon interactions are hidden in the seemingly chaotic remains of the binding energy. The distribution of the remains around zero is examined in greater detail in sect.~\ref{sec fluc}. Also visible are faint substructures around $(60,40)$ and $(90,60)$, which is not seen for any of the calculations. These substructures are discussed in greater detail in sect.~\ref{sec fluc}. The theoretical models do assign individual character to the nuclei, but apparently certain physical aspects are still not adequately described by the models.

To estimate the differences in the remaining fluctuations one possibility is to calculate the rmse values between the results of applying the four point mass relations. Another possibility is to calculate the rmse value between the result of the four point relations and zero. This will indicate the size of the fluctuations which remains, when the smooth aspects are eliminated. This can be done for all extrapolated nuclei, or selectively inside or outside the experimental region. Using eq.~(\ref{eq rmse}) $x_i$ is the result of applying $\Delta_{2n}$ or $\Delta_{2p}$. The reference point $\tilde{x}_i$ can then be any of the other data sets, or the value zero. The results are presented in table \ref{tab d2n overview}.

\begin{table}
\centering
\caption{The comparison between the result of applying a four point structure to the different data sets. The data points $x_i$ can be any of the calculated or the measured binding energies. The reference points $\tilde{x}_i$ are either another data set or zero. When comparing with zero it can either be done inside or outside the experimental region, or for all nuclei. All rmse values are in MeV. \label{tab d2n overview}}
\begin{tabular}{ *{3}{l} *{0}{c} *{1}{r}} 
\toprule
Type            &     Data $(x_i)$       & Reference $(\tilde{x}_i)$ & rmse   \\
\midrule
$\Delta_{2n}$   &     HFB27              &       Exp                 &   0.20 \\
                &     D1SG               &       Exp                 &   0.41 \\
                &     SkP                &       Exp                 &   0.27 \\ \cmidrule{2-4}
                &     HFB27              &       0                   &   0.45 \\
                &     D1SG               &       0                   &   0.52 \\
                &     SkP                &       0                   &   0.55 \\ \cmidrule{2-4}
                &     Exp                &       0 in                &   0.29 \\
                &     HFB27              &       0 in                &   0.36 \\
                &     D1SG               &       0 in                &   0.67 \\
                &     SkP                &       0 in                &   0.39 \\ \cmidrule{2-4}
                &     HFB27              &       0 out               &   0.47 \\
                &     D1SG               &       0 out               &   0.45 \\
                &     SkP                &       0 out               &   0.60 \\ \cmidrule{2-4}
                &     HFB27              &       D1SG                &   0.37 \\
                &     HFB27              &       SkP                 &   0.50 \\
                &     SkP                &       D1SG                &   0.37 \\ \midrule
$\Delta_{2p}$   &     HFB27              &       Exp                 &   0.20 \\
                &     D1SG               &       Exp                 &   0.31 \\
                &     SkP                &       Exp                 &   0.41 \\ \cmidrule{2-4}
                &     HFB27              &       0                   &   0.32 \\
                &     D1SG               &       0                   &   0.46 \\
                &     SkP                &       0                   &   0.56 \\ \cmidrule{2-4}
                &     Exp                &       0 in                &   0.29 \\
                &     HFB27              &       0 in                &   0.29 \\
                &     D1SG               &       0 in                &   0.56 \\
                &     SkP                &       0 in                &   0.49 \\ \cmidrule{2-4}
                &     HFB27              &       0 out               &   0.33 \\
                &     D1SG               &       0 out               &   0.42 \\
                &     SkP                &       0 out               &   0.58 \\ \cmidrule{2-4}
                &     HFB27              &       D1SG                &   0.32 \\
                &     HFB27              &       SkP                 &   0.46 \\
                &     SkP                &       D1SG                &   0.44 \\
\bottomrule
\end{tabular}
\end{table}

Based on the rmse values, the HFB27 model agrees most closely with the measured values. Comparing with table \ref{tab rmse} the rmse for the HFB27 calculation has only decreased by about a factor 3. This rather modest reduction indicates the original deviation between the HFB27 values and the experimental values mostly originated from minor non-systematic variations. On the other hand, the rmse values for both the SkP and the D1SG models are reduced by roughly an order of magnitude. The deviation between these models and the measured values must then be relatively smooth in nature. This corroborates the picture established by figs.~\ref{fig diff exp}b and \ref{fig diff exp}c, where a systematic deviation was visible.

The scale for the rmse of the $\Delta_{2p}$ structure is almost identical. This emphasizes the fact that the conclusions drawn are independent of the specific choice of $n_i$ and $z_i$. As long as those values are small, the second order difference will cancel all smooth aspects up to second order. The same conclusions can then be reached for any small $n_i$ and $z_i$ values, if care is taken to only combine nuclei with either even or odd neutron and proton number.

Since the extrapolations deviate from the experimental values, it is not surprising that they also deviate more from zero than the experimental value does, when applying $\Delta_{2n}$. A slight change can be seen when comparing the result inside and outside the experimental region. For the HFB27 and SkP models the rmse value is slightly larger outside the experimental region, and the opposite is true for the D1SG model. It is not surprising that the HFB27 deviates more from zero outside the experimental region, as very significant fluctuations and substructures are seen for the super-heavy nuclei in both part (a) and part (b) of fig.~\ref{fig D2n ext}. The outlying nuclei fig.~\ref{fig D2n ext}d are also outside the experimental region. In fig.~\ref{fig D2n ext}c the only significant structures are the shells, and they tend to decrease outside the experimental region. A smaller rmse value is then to be expected.

\section{Odd-even effects \label{sec oe}}

Another possibility, when trying to evaluate the accuracy of the individual models, is to examine certain aspects of the binding energy. Here odd-even effects will be isolated, and compared with a phenomenological expression relying on neutron excess.

\subsection{The three point difference \label{sec 3 diff}}

By properly combining neighbouring nuclei, instead of nuclei with even or odd neutron and proton number, it is possible to emphasize specific aspects of the binding energy. A three point structure will be employed. The idea behind this structure is outlined below, but a more detailed explanation can be found in \cite{hov13-2}. The basic principle is the same as the one outlined in sect.~\ref{sec 4 diff}, only a first order difference is used instead
\begin{align}
Q_1(n_1,z_1) =& -B(N-n_1, Z-z_1) +2B(N,Z) \notag \\ &-B(N+n_1,Z+z_1).
\end{align}
The leading order contribution to the smooth part of $Q_1$ is of second order. To further eliminate the smooth parts, a similar combination, $Q_{1-LD}$, only based on the liquid drop model, is calculated and subtracted. A very simple version of the liquid drop model is used
\begin{align}
B_{LD} = a_v A - a_s A^{2/3} - a_c \frac{Z^2}{A^{1/3}}-a_a \frac{(A-2Z)^2}{A},
\end{align}
where $(a_v, a_s, a_c, a_a) = (15.56, 17.23, 0.7, 23.285)$ all in MeV. The final structures used are
\begin{align}
\Delta_n = \frac{1}{2} (-1)^N \left( Q_1(1,0) - Q_{1-LD}(1,0) \right), \notag \\
\Delta_p = \frac{1}{2} (-1)^Z \left( Q_1(0,1) - Q_{1-LD}(0,1) \right).
\end{align}
The purpose of these structures is to emphasize odd-even staggering effects. A similar four point structure could have been created, which would not need to subtract the smooth parts. The three point structure was chosen for its compactness, as a larger structure would be more likely to combine nuclei with distinct, individual character, which would be detrimental to the accuracy of the final result. When only nuclei with either even or odd neutron and proton numbers were combined, structural similarities could more reasonably be assumed, which made compactness less of an issue. With this three point mass relation the measured binding energies have previously been examined \cite{hov13-2}. A decrease of the odd-even effects with neutron excess was found.

\subsection{Application of the three point difference}

A necessary requirement for the three point mass relations to be applicable is that the calculation contains both even and odd nuclei. This is only the case for the HFB27 calculation. However, included in the SkP data set are explicitly calculated $\delta_n$ and $\delta_p$ pairing gaps, which should account for part of the odd-even effects with respect to neutrons and protons.

\begin{figure}
\centering
\includegraphics[width=1.0\linewidth]{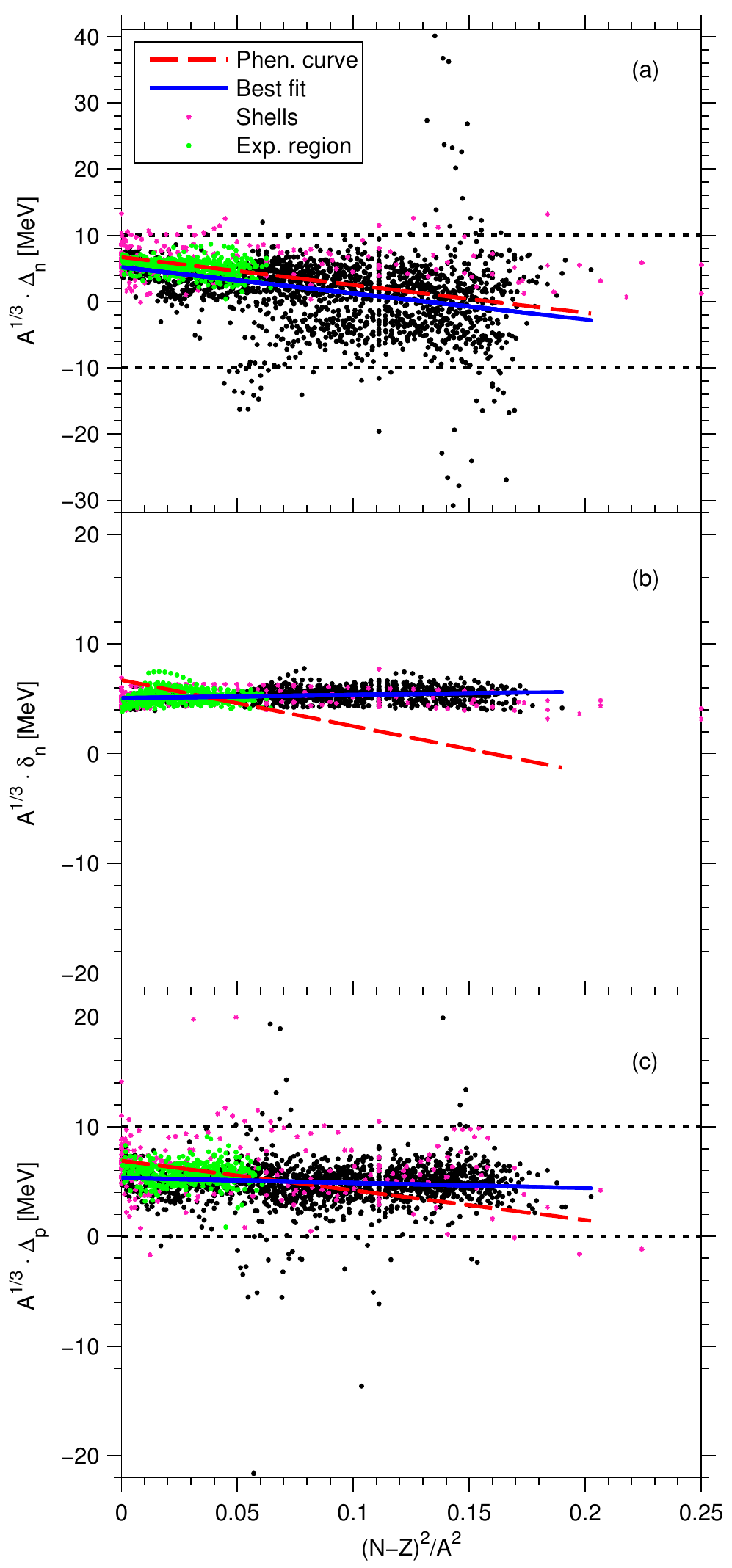}

\caption{(Colour online) (a) The three point $\Delta_{n}$ mass relation applied to the even-even HFB27 calculations as a function of neutron excess, (b) the $\delta_n$ values from the SkP data set presented similarly, and (c) the three point $\Delta_{p}$ mass relation applied to the even-even HFB27 calculations. The nuclei marked in green lie inside the experimental region, the nuclei marked in purple are influenced by either shell or Wigner effects, the dashed, red line is a phenomenological curve \cite{hov13-2}, and the blue line is the best linear fit. Nuclei outside the dotted lines are shown in Fig.~\ref{fig out G}. \label{fig pair ee}}
\end{figure}

The result of applying the three point structure to all even-even nuclei in the HFB27 calculation, scaled by $A^{1/3}$, as a function of $(N-Z)^2/A^2$, is seen in fig.~\ref{fig pair ee}a. Marked in green are the nuclei inside the experimental region, and marked in light blue are nuclei which are expected to deviate from the rest. These are either influenced by shell structures or by the Wigner effect. The red curve is a phenomenological expression for the $N-Z$ dependence of the odd-even effects \cite{hov13-2}. It is the result of a least squares fit of $A^{1/3} \Delta_{n}$ as a function of $(N-Z)^2/A^2$ for the measured binding energies. The blue curve is the best linear fit to all calculated even-even nuclei. The linear decrease seen in the experimental region is continued even for the super heavy, although the scattering increases greatly. It should be noted that in the model for the HFB27 data set, the pairing forces do not contain an explicit neutron excess dependence \cite{gor10}. The slight neutron excess dependence must be an indirect consequence of the particular mean-field calculation with different pairing strengths.

In fig.~\ref{fig pair ee}b the $\delta_n$ values from the SkP data set are scaled with $A^{1/3}$ and presented as a function of neutron excess, similar to what was seen in fig.~\ref{fig pair ee}a. Again nuclei marked in green are inside the experimental region, light blue are influenced by shells or the Wigner effect, and the red curve is the same phenomenological expression as in fig.~\ref{fig pair ee}a. The blue curve is the best linear fit to all nuclei. The result is almost constant both inside and outside the experimental region, and shows no dependence on neutron excess. Here the nuclei marked in blue do not tend to deviate from the rest. There is no $N-Z$ dependence, as the total pairing gap is a constant and the gab has only marginal structure originally from the mean-field spectrum.

The equivalent to fig.~\ref{fig pair ee}a using the three point proton structure, $\Delta_p$, is seen in fig.~\ref{fig pair ee}c. It is worth noticing that not only is the scattering significantly less than for the neutron structure, the result is also almost constant. The mean field proton energy levels are less affected by adding another nucleon when inside the Coulomb barrier. A significantly less pronounced neutron excess dependence is then emerging, in particular for the very proton rich nuclei.

\begin{figure}
\centering
\includegraphics[width=1.0\linewidth]{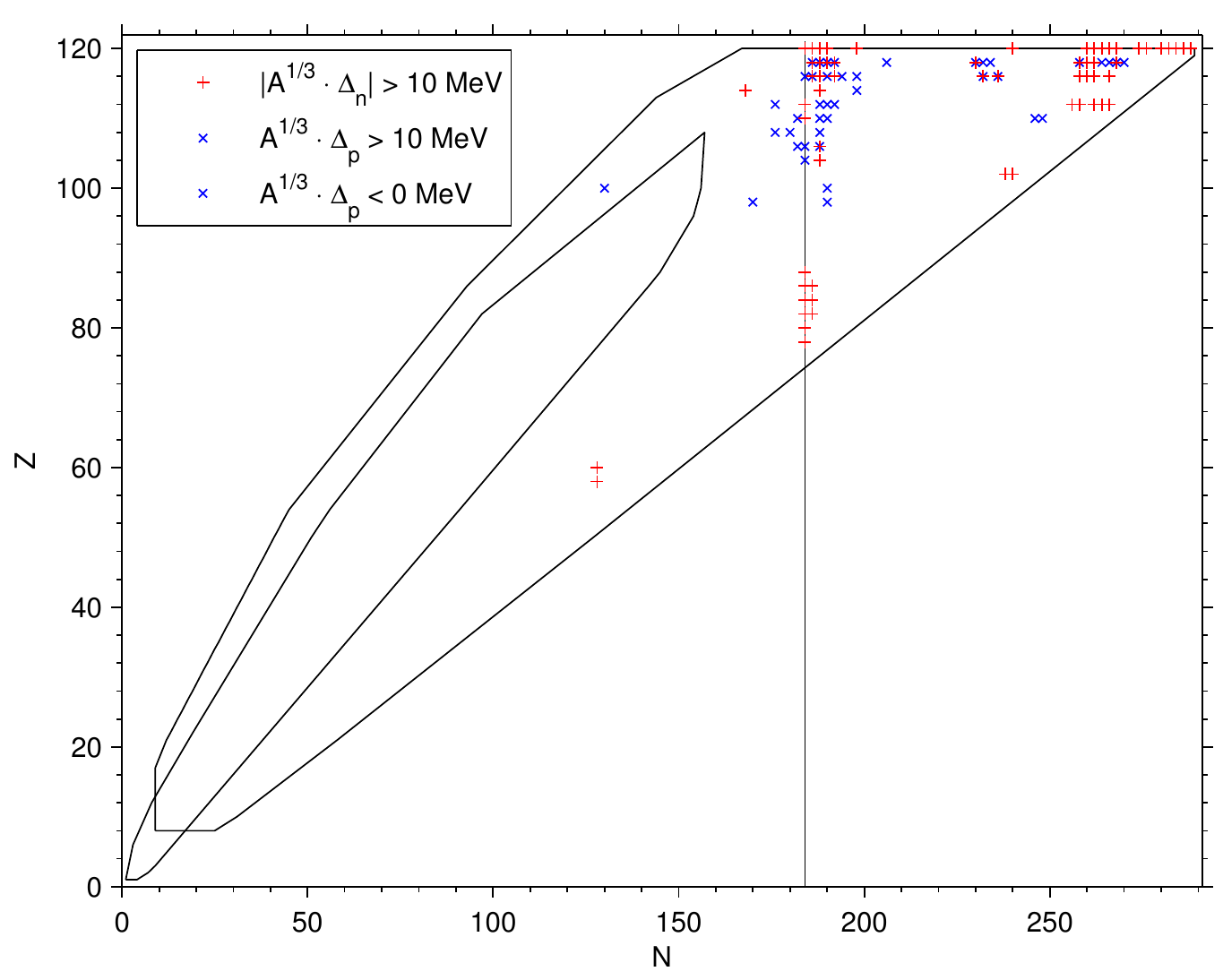} 
\caption{(Colour online) The nuclei from fig.~\ref{fig pair ee}a which lies outside $\pm 10 \, \si{\mega\electronvolt}$ and from fig.~\ref{fig pair ee}c which lies above $+10 \, \si{\mega\electronvolt}$ or below $0 \, \si{\mega\electronvolt}$. Outlined is the experimental region and the HFB27 data set. The vertical line is at $N=184$. \label{fig out G} }
\end{figure}

The nuclei from figs.~\ref{fig pair ee}a and \ref{fig pair ee}c, where $\lvert A^{1/3} \Delta \rvert > 10 \, \si{\mega\electronvolt}$ are shown in the chart of nuclei in fig.~\ref{fig out G}. Included are the outline of the experimental region and the outline of the HFB27 data set. The suspected shell at $N = 184$ is also shown. Most of the outliers group around this line, or in the region which could contain an island of stability. It is then not surprising these nuclei deviate substantially more than the other.

The result of applying $\Delta$ (not scaled with $A^{1/3}$) to the calculations and to the experimental values is compared in table \ref{tab pair rmse}. This is done for both neutrons and protons for all nuclei (when available) as well as for even-even nuclei only. The rmse$(\Delta)$ value is less than $0.3 \, \si{\mega\electronvolt}$, but it is slightly worse than the $0.2 \, \si{\mega\electronvolt}$ found in table \ref{tab d2n overview}. It should be noted that, in spite of its simplicity, $\delta_n$ is actually closest to the experimental value. It follows that the odd-even mass effects not included in the pairing gap are minor corrections.

\begin{table}
\centering
\caption{Comparison between the odd-even effects for both neutrons and protons. Either all nuclei or only even-even nuclei are included. The rmse value in the third column is between the result in MeV of $\Delta_{n/p}$ (or $\delta_{n,p}$) for the calculations and the experimental measurements. \label{tab pair rmse}}
\begin{tabular}{ l *{2}{c}} 
\toprule
Type        &   Nuclei      &   rmse($\Delta$)   \\
\midrule
$\Delta_n$  &   e-e         &      0.25          \\
            &   All         &      0.29          \\
$\delta_n$  &   e-e         &      0.22          \\
$\Delta_p$  &   e-e         &      0.24          \\
            &   All         &      0.26          \\
$\delta_p$  &   e-e         &      0.52          \\
\bottomrule
\end{tabular}
\end{table}

Another possibility is to compare the results to the phenomenological curves seen in fig.~\ref{fig pair ee}. The result for even-even nuclei is seen in table \ref{tab pair our}, where the rmse values are calculated as in table \ref{tab pair rmse}. The HFB27 calculations and the $\delta$ values are compared to the phenomenological curve (the red curve), and to the best global fit (the blue curve). The experimental results are only compared to the red phenomenological curve, as this is the best fit inside the experimental region.

\begin{table}
\centering
\caption{Comparing the extrapolations and the experimental results for odd-even effects to the phenomenological curves from fig.~\ref{fig pair ee}. This is done for both the neutron and proton structure, as indicated by the first column. The second column specifies which data set is compared to the phenomenological curve, while the third column specifies whether all nuclei, only those inside, or those outside the experimental region are being used. The fourth column indicates which of the curves from fig.~\ref{fig pair ee} are being compared to; this is either the best fit to all nuclei (blue curve in fig.~\ref{fig pair ee}) or the (red, dashed) phenomenological curve, which is a fit to the experimental measurements. The final column is the rmse values calculated as in table \ref{tab pair rmse}. \label{tab pair our}}
\begin{tabular}{ l l l l *{1}{c}} 
\toprule
   Type     &   Data      &   Region   &   Curve        &   rmse($\Delta$)   \\
\midrule
$\Delta_n$  &   HFB27     &   All      &    Fit exp     &      0.74          \\
            &   HFB27     &   Out. Exp &    Fit exp     &      0.84          \\
            &   HFB27     &   In. Exp  &    Fit exp     &      0.23          \\
            &   HFB27     &   All      &    Best fit    &      0.73          \\
            &   Exp       &   In. Exp  &    Fit exp     &      0.17          \\
$\delta_n$  &   SkP       &   All      &    Fit exp     &      0.47          \\
            &   SkP       &   Out. Exp &    Fit exp     &      0.57          \\
            &   SkP       &   In. Exp  &    Fit exp     &      0.21          \\
            &   SkP       &   All      &    Best fit    &      0.11          \\
$\Delta_p$  &   HFB27     &   All      &    Fit exp     &      0.35          \\
            &   HFB27     &   Out. Exp &    Fit exp     &      0.39          \\
            &   HFB27     &   In. Exp  &    Fit exp     &      0.21          \\
            &   HFB27     &   All      &    Best fit    &      0.30          \\
            &   Exp       &   In. Exp  &    Fit exp     &      0.15          \\
$\delta_p$  &   SkP       &   All      &    Fit exp     &      0.38          \\
            &   SkP       &   Out. Exp &    Fit exp     &      0.27          \\
            &   SkP       &   In. Exp  &    Fit exp     &      0.52          \\
            &   SkP       &   All      &    Best fit    &      0.09          \\
\bottomrule
\end{tabular}
\end{table}

Once again the rmse value of the HFB27 calculation is in the $0.2-0.3 \, \si{\mega\electronvolt}$ range inside the experimental region. Naturally, there is a closer agreement with the experimental data, as the curve is obtained as a least squares fit to those. For the neutron structure the error value is significantly larger outside the experimental region, because of the large scattering seen in fig.~\ref{fig pair ee}a. As a result of less scattering, the proton structure has a much better agreement with the phenomenological curve outside the experimental region than the neutron structure.

\section{Random fluctuations \label{sec fluc}}

\begin{figure}
\centering
\subfloat[]{\includegraphics[width=1.0\linewidth]{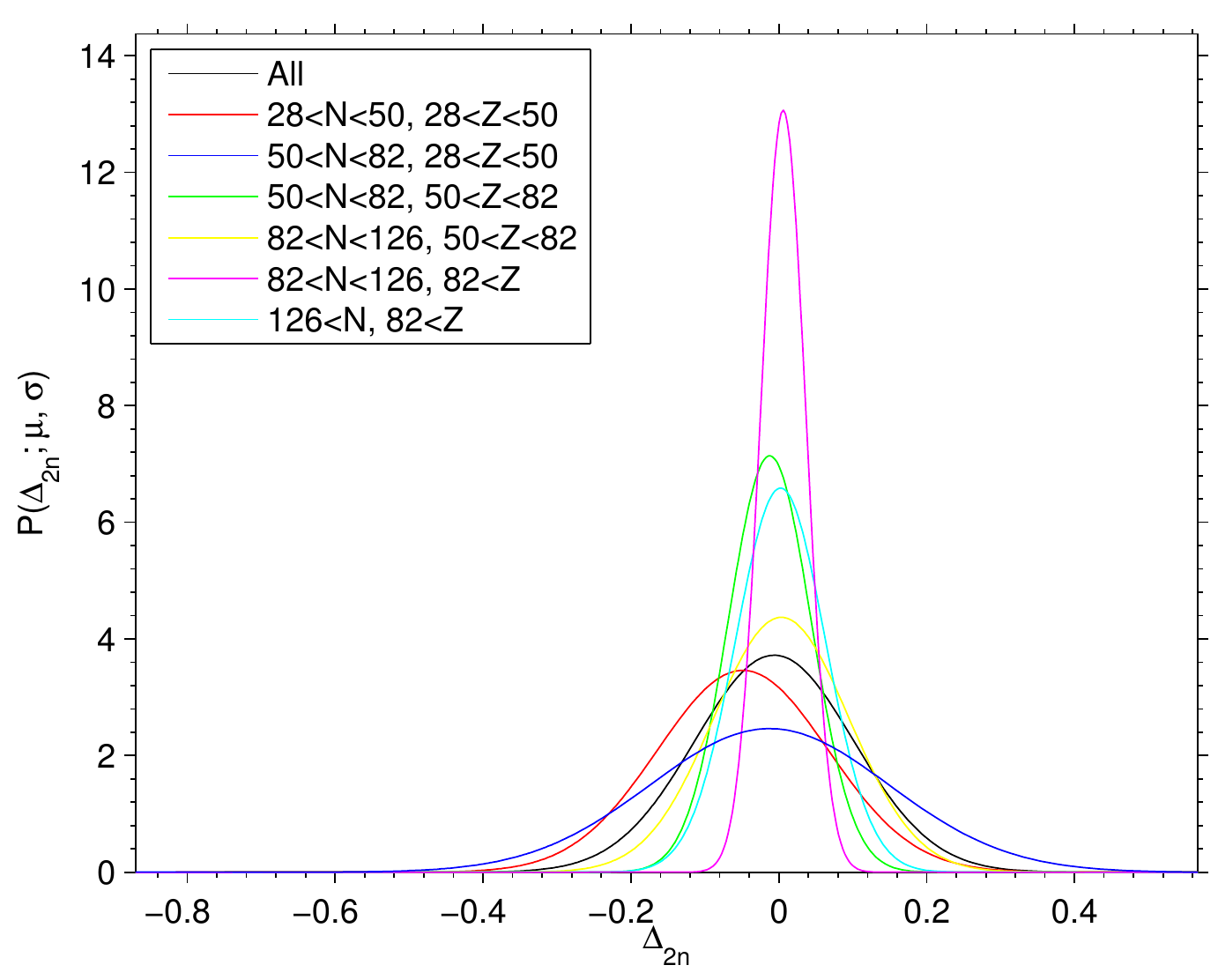} \label{fig gau exp}}

\subfloat[]{\includegraphics[width=1.0\linewidth]{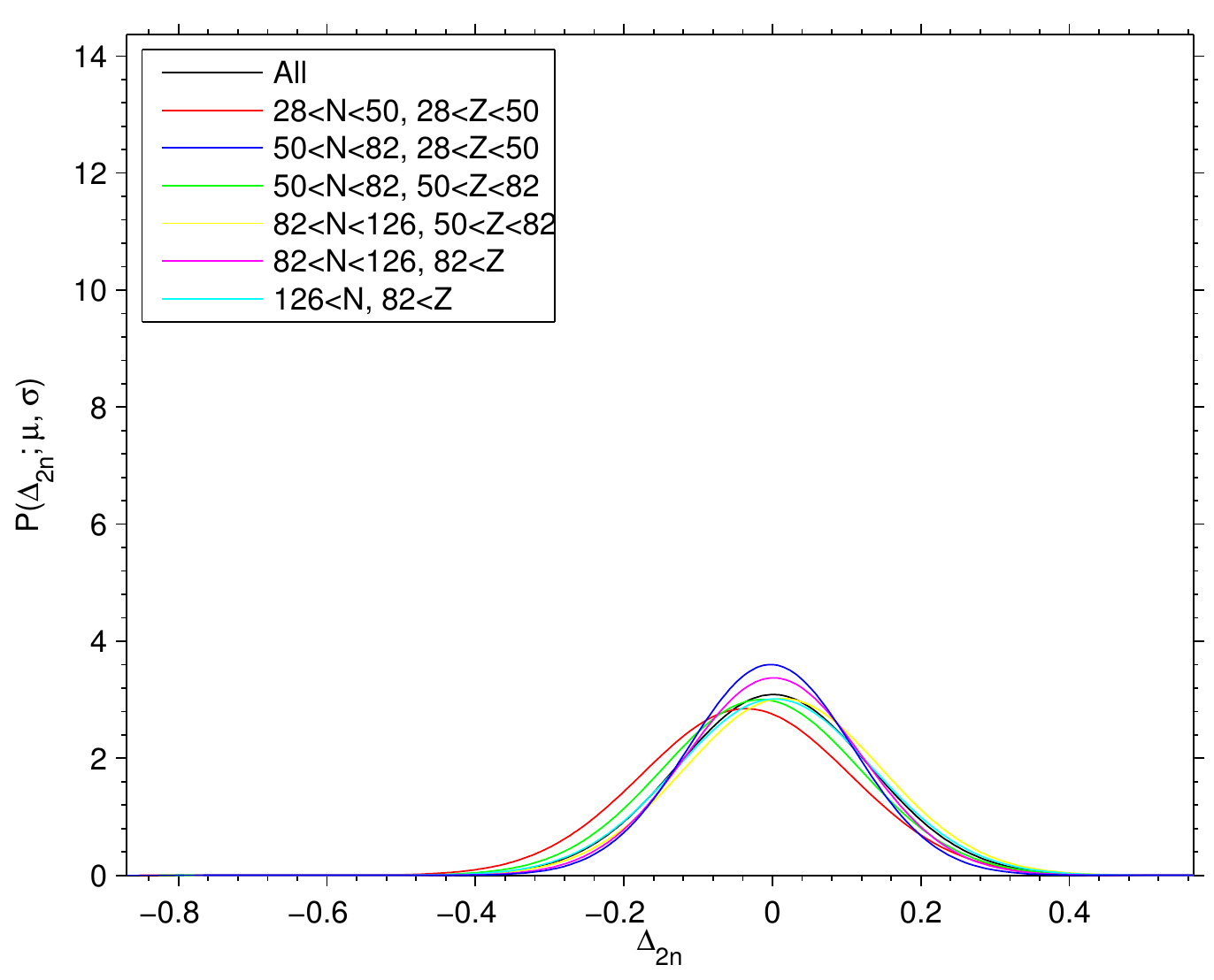} \label{fig gau gor}}
\caption{(Colour online) The Gaussian distributions for the result of applying $\Delta_{2n}$ in the regions outlined by the known shells. (a) For the measured binding energies, and (b) for the HFB27 calcualtions. \label{fig gau}}
\end{figure}

Because of erratic fluctuations mass relations, such as $\Delta_{2n}$, do not cancel completely, as seen in sect.~\ref{sec 4 diff}. The present section will focus on the distribution of what remains, to detect structural differences. The remains are distributed closely around zero. If nuclei including known effects, such as shells or the Wigner effect, are neglected, the distribution is reasonably well described by a Gaussian distribution 
\begin{align}
P(x; \mu, \sigma) = \frac{1}{\sigma \sqrt{2 \pi}} \exp\left( -\frac{1}{2} \left( \frac{x-\mu}{\sigma} \right)^2 \right),
\end{align}
where $\mu$ and $\sigma$ are the mean and the standard deviation respectively. Figure \ref{fig gau exp} contains the Gaussian distributions of $\Delta_{2n}$ applied to the measured binding energies between each of the known major shells, as well as the distribution for all nuclei. The same, only for the HFB27 calculation is seen in fig.~\ref{fig gau gor}. The mean and the standard deviation for these distributions are included in table \ref{tab gau}. The D1Sg and SkP calculations contained too few nuclei for such evaluations to be sensible.

The theoretical Gaussian distributions in fig.~\ref{fig gau gor} have a larger width as reflection of the fact that the computed masses on average deviate more than the measured values. To understand this it is worth recapitulating that the $\Delta_{2n}$ mass relation is a measure of the local smoothness of the masses in the nuclear chart. The mean-field approximation is then fundamentally very sensitive to the shell structure around the Fermi energy because adding one nucleon necessarily occupies the next level wherever it is. This sharp structure is smeared out by the adjustments from the self consistency condition and the included pairing correlations. Other correlations are not accounted for in these models. On the other hand, we observe that the experimental masses are smoother since all possible correlations may contribute.

\begin{table*}
\centering
\caption{The Gaussian distributions of $\Delta_{2n}$ in the regions given by the known shells. The mean and the standard deviation for both the experimental measurements and the HFB27 calculations are included, along with the number of particles in each region. The third column specfies the number of nuclei in each region. \label{tab gau}}
\begin{tabular}{ *{3}{c} S[table-format=2.7] S[table-format=1.7] S[table-format=2.7] S[table-format=1.7] } 
\toprule
\multicolumn{2}{c}{Region}&    &  \multicolumn{2}{c}{Exp.} & \multicolumn{2}{c}{HFB27} \\
\cmidrule(lr){1-2} \cmidrule(lr){4-5} \cmidrule(lr){6-7}
$N$    & $Z$   & \# &  \multicolumn{1}{c}{$\mu$}  & \multicolumn{1}{c}{$\sigma$} & \multicolumn{1}{c}{$\mu$}  & \multicolumn{1}{c}{$\sigma$}\\
\midrule
28-50  & 28-50 &    55  &   -0.05(2)   &   0.115(11)   &   -0.04(2)   &   0.140(13) \\
50-82  & 28-50 &    191 &   -0.012(12) &   0.162(8)    &   -0.03(8)   &   0.111(6)  \\
50-82  & 50-82 &     93 &   -0.012(6)  &   0.056(4)    &   -0.016(14) &   0.133(10) \\
82-126 & 50-82 &    437 &    0.004(4)  &   0.091(3)    &   -0.014(6)  &   0.132(5)  \\
82-126 & 82-   &     38 &    0.006(5)  &   0.031(4)    &    0.002(19) &   0.118(14) \\
126-   & 82-   &    160 &    0.002(5)  &   0.061(3)    &    0.003(11) &   0.132(7)  \\
All    & All   &   1003 &   -0.006(3)  &   0.107(2)    &    0.001(4)  &   0.129(3)  \\
\bottomrule
\end{tabular}
\end{table*}

\begin{figure}
\centering
\includegraphics[width=1.0\linewidth]{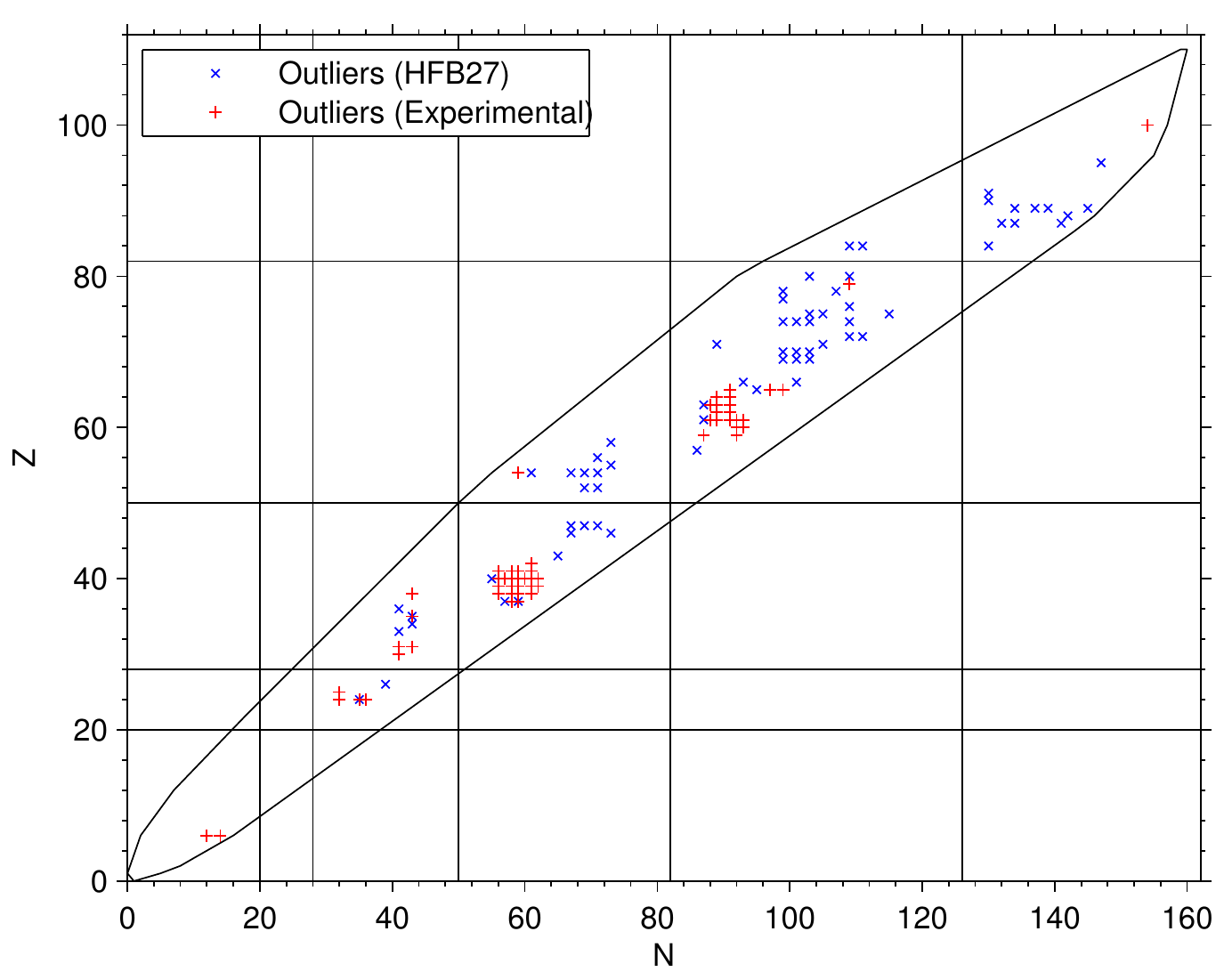} 
\caption{(Colour online) The nuclei more than $2\sigma$ away from $\mu$ in fig.~\ref{fig gau}. Outliers for both figs.~\ref{fig gau exp} and \ref{fig gau gor} are included. \label{fig 2sig out}}
\end{figure}

By considering the nuclei which deviate significantly from the mean of the distribution, minor substructures can be detected. In fig.~\ref{fig 2sig out} the nuclei, which lie more than two standard deviations from the center of the Gaussian distribution of all nuclei, are highlighted. The outliers in the experimental data set show a tendency to group around $N=60$ and $Z=40$ as well as around $N=90$ and $Z=60$. These are the same substructures as was faintly seen in fig.~\ref{fig D2n E}. These regions are subject to a very rapid change in deformation \cite{hey11}. Localized changes in the curvature of the mass surface would be detected by a second order difference, and might well contribute to the observed structures. As such these substructures reflect, at least partly, a systematic effect, and cannot be considered part of the binding energy's possibly chaotic fluctuations. If these structures were omitted, there would likely be a clear decrease in scattering from light to heavy nuclei.

The outliers in the extrapolated data set are much more evenly distributed between the shells. There are no signs of groupings, or of a decrease with nucleon number. This seemingly chaotic distribution of outliers once again demonstrates the impressive individual character of the nuclei in  the HFB27 calculations. Unfortunately, these specific, localized substructures are not reproduced in the global model. Here outlying nuclei were defined as being more than $2 \sigma$ away from $\mu$. The conclusions do not depend on this relatively arbitrary distance. There is always a grouping of the experimental outliers, and never a grouping of the calculated HFB27 outliers, independent of the distance defining these outliers.

\section{Summary and conclusion}

The purpose of this paper is to asses the accuracy of three sets of calculated binding energies, as well as to determine the structural differences between measured and calculated values. Three different Hartree-Fock-Bogoliubov models are considered. The first is a rather elaborate model (HFB27), designed with the explicit purpose of calculating nuclear mass excess. The second is a simpler model (SkP), which is computationally much less demanding. The third is a slightly different model (D1SG), whose primary focus is not the calculation of binding energies. We begin by considering a simple difference between the data sets. This included both the difference between the calculated and the measured binding energies, and the differences between each of the three calculations.

To highlight the minor, non-smooth variations a second order, four-point mass relation was applied. This mass relation is rather flexible, and several different nuclei combinations are possible. Common to all is the fact that all smooth aspects of the binding energy are eliminated to second order. All odd-even effects are also eliminated by choosing a mass relation combining only either even or odd neutron and proton numbers. The primary mass relation, $\Delta_{2n}$, is horizontal in the $(N,Z)$ plane, combining only one specific proton number and four neighbouring neutron numbers, that is $N, \, N\pm 2$ and $N+4$. To demonstrate that the results did not rely on the specific configuration of the mass relation, a similar proton structure, $\Delta_{2p}$, was also applied.

Specific aspects of the binding energy were then considered more closely. By introducing a three-point mass relation, which combined neighbouring nuclei, odd-even effects could be isolated. This is only possible for the HFB27 calculations, as both the D1SG and the SkP calculations only include even-even nuclei. Previous work using this mass relation led to a phenomenological expression for the odd-even effects as a function of neutron excess. This expression was included and extended outside the experimentally known region. The $\delta_n$ and $\delta_p$ values included in the SkP data set, which described the pairing gaps, were treated similarly. 

The remains after applying the $\Delta_{2n}$ mass relation were then examined in detail. A Gaussian distribution was calculated in each region, outlined by the known shells. Too few nuclei were calculated using the D1SG and the SkP calculations to allow for such analysis. Only for the HFB27 calculations and the measurements could the Gaussian distributions be compared. By considering the statistical outliers in the distributions, structural differences could also be detected.

The simple, initial differences very clearly demonstrates the existence of disparities between calculated and measured binding energies. A systematic deviation, increasing with distance to stability, is seen for the D1SG calculations, with the calculations being too small. For the SkP calculations the deviation increases roughly with proton number, whereas the deviation of the HFB27 calculation is both less pronounced and more erratic. The overall scale of the deviation is indicated by the root-mean-square-error values, which are around $4.7$ and $3.2$ MeV for the D1SG and and SkP calculations respectively, and only around $0.7$ MeV for the HFB27 calculation.

Applying $\Delta_{2n}$ and $\Delta_{2p}$ resulted in a very unobstructed view of the fluctuations in the binding energy. The calculated binding energies fluctuated significantly more than the measured binding energies, in particular for the elaborate HFB27 model. This also faintly revealed some interesting substructures in the measured binding energies around $(N,Z) = (56-62,37-41)$ and $(88-93,60-64)$. The scale of the remains of these substructures, after applying the mass relations, is only around $0.5$ MeV. For comparison the scale of the remains of the shells is around $1$ MeV. There are several possible explanations for these substructures. Both are regions with a rapid onset of deformation, and the substructures are also on supposed minor shells. However, without knowing the exact cause of these deviations, similar deviations of the same magnitude are possible elsewhere.

When isolating the odd-even effects using $\Delta_n$ a decrease as a function of neutron excess is clearly seen. There are significant scattering, but the outliers tend to group around $N=184-186$, which has all the characteristics of a major shell. It should be noted that this neutron excess dependence comes out of the calculations without being explicitly included in the model. The HFB27 calculation contain four pairing parameters, none of which include an explicit $N-Z$ dependence. When applying $\Delta_p$ the result is essentially constant. A possible explanation is that the larger Coulomb barrier of the heavier nuclei effectively traps the protons. The $\delta_n$ and $\delta_p$ values are estimates of the total pairing gap, and did not decrease with neutron excess. In calculating $\delta$ it is assumed that an odd nucleus is just an even nucleus with an extra nucleon. With only two pairing parameters and such a simple model it is then difficult to incorporate more elaborate dependencies.

Applying the $\Delta_{2n}$ mass relation resulted in a Gaussian distribution around zero, reflecting the fact that the mass relation very effectively eliminates everything but the minor erratic fluctuations. The Gaussian distributions for the measured values show significant variation between regions. Also a tendency for the scattering to decrease with nucleon number is observed, although local substructures affected this trend. The distributions of the HFB27 calculations do not exhibit any regional changes. The scattering, for the HFB27 calculations in all regions, is comparable to the largest scattering of the measured values. The outlying nuclei, in the tails of the distributions, reveal very distinct, structural differences. The experimental outliers clearly either group around $(N,Z) = (60,40)$ and $(90,60)$, or are found among the light nuclei. Both these regions are known for rapidly changing deformations, and it has been suggested that both 40 and 64 are minor magic numbers. The outliers in the HFB27 calculation are scattered throughout the chart of nuclei, without any discernible patterns or groupings.

In summary, we compared three different Hartree-Fock-Bogoliubov calculations of nuclear binding energies to the measured values. Using specifically designed, second order, four-point mass relations the minor fluctuations in the binding energies are isolated. This reveals small scale structural differences between the calculated and the measured binding energies. By a similar three-point mass relation the odd-even effects are isolated. An unintended decrease with neutron excess is found for the calculated binding energies even well outside the experimentally known region. 

We have shown some of the limitations of current mean-field calculations. These are small scale limitations, which demonstrate the overall accuracy of refined Hartree-Fock-Bogoliubov calculations. However, as seen from fig.~\ref{fig gau} the experimental masses are clearly more correlated (over mass numbers differing by 6 units) than the theoretical models currently predict.

This work was funded by the Danish Council for Independent Research DFF Natural Science and the DFF Sapere Aude program.

\end{document}